\documentclass[%
 reprint,
superscriptaddress,
nofootinbib,
 amsmath,amssymb,
 aps,
prd
]{revtex4-2}

\usepackage{graphicx}
\usepackage{dcolumn}
\usepackage{bm}
\usepackage{hyperref}
\usepackage[usenames,dvipsnames]{xcolor}
\hypersetup{
    colorlinks = true,
    citecolor = {MidnightBlue},
    linkcolor = {BrickRed},
    urlcolor = {BrickRed}
}

\usepackage{aas_macros}

\newcommand{\F}{\mathcal{F}}	
\newcommand{\G}{\mathcal{G}}	
\newcommand{\Pow}{\mathcal{P}}	

\begin{document}

\title{Weak gravitational flexion in various spacetimes:\\ Exotic lenses and modified gravity}

\author{Evan J. Arena}
  \email{evan.james.arena@drexel.edu}
  \homepage{http://evanjarena.github.io}
 \affiliation{%
Department of Physics, Drexel University, Philadelphia, PA 19104, USA
}

\date{\today}

\begin{abstract}
Exotic objects such as the Ellis wormhole are expected to act as gravitational lenses.  Much like their nonexotic counterparts, information about these lenses can be found by considering the strong and weak lensing fields they induce.  In this work, we consider how weak gravitational lensing flexion can provide information beyond that of shear.  We find that directional flexion can distinguish between the case of a positive or negative convergence where directional shear cannot, and therefore can provide a unique lensing signature for objects with an Ellis wormhole-type metric.  We also consider cosmic flexion, the flexion correlation function whose signal originates from the large-scale structure of the Universe, in the context of modified gravity.  We find flexion to be a unique probe of parametric models of modified gravity, particularly in the case of scale-dependent phenomenological post-general relativity functions.

\end{abstract}

\maketitle


\section{Introduction}

Gravitational lensing has become one of the quintessential cosmological and astrophysical probes of the last few decades \citep{Bartelmann:1999yn, Kilbinger:2014cea, DES:2021vln}. Lensing probes the gravitational potential and is therefore a useful measure of the total matter distribution.  To this end, lensing has had a great impact at several different mass scales. Lensing is powerful for studying galaxy cluster mass distributions \citep{1987A&A...172L..14S, 1990ApJ...349L...1T}.  A weaker effect, known as galaxy-galaxy lensing, is the lensing of a background galaxy by a foreground galaxy. Specifically, galaxy-galaxy shear correlates the shapes of high-redshift ``source'' galaxies with positions of low-redshift ``lensing'' galaxies \citep{Brainerd:1995da, SDSS:1999zww}.  Even weaker is the lensing by the large-scale structure of the Universe -- specifically, the so-called cosmic shear -- which probes the underlying matter power spectrum \citep{Bacon:2000sy, Kaiser:2000if, vanWaerbeke:2000rm, Wittman:2000tc}. Finally, lensing of the cosmic microwave background (CMB) has also been detected at high significance \cite{Aghanim:2018eyx}, which has been a further useful probe of cosmological parameters. 

In weak lensing, there exist a variety of simple lens models that make the assumption of a circularly symmetric profile.  These include the Schwarzschild lens (see e.g. Ref. \cite{Bartelmann:1999yn}), the singular isothermal sphere (SIS; see e.g. Ref. \cite{1987gady.book.....B}), and the Navarro-Frenk-White (NFW) density profile \citep{Navarro:1994hi, Navarro:1995iw, Navarro:1996gj}, to name a few. Recently, exotic lens models have been discussed in which the lens can can repulse rather than attract light, by having, for example, negative mass.  These models have been inspired by modified gravity theories as well as individual exotic objects such as the Ellis wormhole, which is a particular example of the Morris-Thorne traversable wormhole class \citep{Ellis:1973yv, Morris:1988cz, Morris:1988tu}. In Ref. \cite{Kitamura:2012zy}, a family of exotic lens models is constructed by considering an exotic spacetime metric that is static and circularly symmetric, can describe both positive and negative masses, and depends on the inverse distance to the power of $n$ (e.g. $n=1$ for Schwarzschild metric, $n=2$ for Ellis wormhole). Then, in Ref. \cite{Izumi:2013tya}, the gravitational shear is worked out for these exotic lenses, and it is shown that the shear can exhibit behavior that suggests a positive mass lens in the presence of an exotic lens.
 
In cosmology, studies of the CMB have found that observations agree with the standard cosmological model ($\Lambda$CDM) to remarkable accuracy \citep{Blumenthal:1984bp, SupernovaSearchTeam:1998fmf, SupernovaCosmologyProject:1998vns, Aghanim:2018eyx}. As we look at more recent parts of cosmic history, using tools such as weak lensing, $\Lambda$CDM still appears to be the law of the land. Subtle discrepancies are found, however,  between these low-redshift observations and the high-redshift measurements of the CMB \cite{DES:2021wwk, Aghanim:2018eyx, Riess:2020fzl, Huang:2019yhh}. These discrepancies could indicate that $\Lambda$CDM might not be sufficient to connect all parts of the cosmic history \citep{Frusciante:2019xia}.  It is therefore necessary to have multiple cosmological probes that complement each other in order to get the full picture of cosmology across all length scales and cosmic times.  To this end, the effort to explain the origin of cosmic acceleration has uncovered a vast zoo of dark energy and modified gravity models. These can be broadly classified according to how they modify general relativity (GR) or replace the cosmological constant, $\Lambda$ -- for example, by adding new scalar, vector or tensor fields (e.g. the broad class of Horndeski models \cite{Bloomfield:2012ff, Bellini:2014fua}); adding extra spatial dimensions; introducing higher-derivative or nonlocal operators in the action; or introducing exotic mechanisms for mediating gravitational interactions \cite{JaiKho10,Clifton:2011jh, Weinberg13, Joyce15, Joyce16, Amendola18}. A systematic study of these models suggests a
number of new gravitational phenomena that can arise if
there are any deviations from the standard cosmological model. These
include the possibility of time- and scale-dependent variations in the gravitational constant, leading to
modifications to the growth rate of large-scale structure and gravitational lensing \citep{2013LRR....16....6A, 2013arXiv1309.5389J, 2013PhRvD..87b3501A, 2014PhRvD..90l4030B, 2015PhRvD..91h3504L}.  These deviations from how gravitational lensing behaves in GR can be studied in the context of strong lensing (see Ref. \cite{Collett:2018gpf}) and with cosmic shear by employing both parametric and nonparametric approaches \citep{2001PhRvD..64h3004U, 2006PhRvD..74b3512K, 2007arXiv0712.1599D, 2008JCAP...04..013A, 2009MNRAS.395..197T, 2009PhRvD..80b3532D, 2010PhRvD..81h3534B, 2010PhRvD..81j3510Z, 2010PhRvD..82j3523D, 2011A&A...530A..68T, 2015MNRAS.454.2722H}.

Beyond shear, there exists a higher-order lensing effect known as \textit{flexion} \citep{Goldberg:2004hh, Bacon:2005qr, Okura:2007js, Schneider:2007ks}.  Like shear, flexion occurs in the presence of single lenses.  There also exists cosmic flexion which, like cosmic shear, is a correlation function whose signal originates from the large-scale structure of the Universe \citep{Bacon:2005qr, AGB}.  The goal of this work is to consider how weak gravitational flexion behaves in the context of two different types of spacetime metrics. In Sec. \ref{sec:2} and \ref{sec:3}, we consider flexion in the presence of a family of exotic lenses, and in Sec. \ref{sec:4} we consider how cosmic flexion behaves in a parametrized modified gravity model. We conclude in Sec. \ref{sec:con}.

\section{Gravitational lensing in single-lens metrics}{\label{sec:2}}

\subsection{The Schwarzschild metric}

Discussions of gravitational lensing typically begin by considering the simple case of light deflection by a point mass $M$.  We start with the Schwarzschild metric \cite{1916SPAW.......189S}
\begin{equation}
    g_{\mu\nu} = 
    \begin{pmatrix}
    -(1-R_s/r) & 0 & 0 & 0\\
    0 & (1-R_s/r)^{-1} & 0 & 0\\
    0 & 0 & r^2 & 0\\
    0 & 0 & 0 & r^2\sin^2\theta 
    \end{pmatrix}
\end{equation}

\noindent where we are using units $c=G=1$, and $R_s = 2M$ is the Schwarzschild radius. In the weak-field (i.e. far-field) limit $r \gg R_s$, the line element of the metric ($ds^2 = g_{\mu\nu}dx^\mu dx^\nu$) simply becomes
\begin{equation}
    ds^2 \simeq -\left(1 - \frac{R_s}{r}\right)dt^2 + \left(1+\frac{R_s}{r}\right)dr^2  + r^2d\Omega^2,
\end{equation}

\noindent where $d\Omega^2 \equiv d\theta^2 + \sin^2\theta d\phi^2$. Since this spacetime is spherically symmetric, we can consider only the equatorial plane $\theta = \pi/2$ without loss of generality, and the deflection angle of light is simply \cite{1992grle.book.....S}
\begin{equation}\label{eqn:deflection_sch}
    \hat{\alpha} = \frac{2R_s}{b}\int_0^{\pi/2}d\phi\cos\phi = \frac{2R_s}{b}
\end{equation}

\noindent where $b$ is the impact parameter. 

In the thin lens approximation, the lensing equation describes the coordinate mapping from the foreground (angular position of the image relative to the lens), ${\bm \theta}$\footnote{Not to be confused with the coordinate $\theta$ in the Schwarzschild metric}, to background (angular position of the source), ${\bm \beta}$, positions via \cite{Bartelmann:1999yn}
\begin{equation}\label{eqn:lens_eqn}
    \bm{\beta} = \frac{\bm{b}}{D_{\rm d}} - \frac{D_{\rm ds}}{D_{\rm s}}\hat{\bm{\alpha}}(\bm{b}) \equiv \bm{\theta} - \bm{\alpha}(\bm{\theta}),
\end{equation}

\noindent where $D_{\rm d}$, $D_{\rm s}$, and $D_{\rm ds}$ are the observer-lens, observer-source, and lens-source distances, respectively, $\bm{\theta} \equiv \bm{b}/D_{\rm d}$, and the reduced deflection angle is defined as $\bm{\alpha} = \hat{\bm{\alpha}} D_{\rm ds}/D_{\rm s}$. 

\subsection{Weak lensing formalism}

In the thin-screen approximation, one defines the convergence, which is related to a dimensionless lensing potential, $\psi$, with $\nabla^2 \psi = 2\kappa$. The lensing potential is the two-dimensional analogue of the Newtonian gravitational potential, integrated along the line of sight. Convergence is a key lensing quantity, which can be simultaneously thought of as a projected, dimensionless surface-mass density of matter, and as an isotropic increase or decrease of the observed size of a source image. In the weak lensing regime, we can rewrite the lens equation as
\begin{equation}
    \beta_i = \delta_{ij}\theta^j - \psi,_{ij}\theta^j - \frac{1}{2}\psi,_{ijk}\theta^j\theta^k
\end{equation} 

\noindent (where $\psi,_{ij}$ is shorthand for $\partial_i\partial_j\psi$ and $\partial_i = \partial/\partial\theta_i$).  We define a complex gradient operator $\partial = \partial_1+i\partial_2$, such that 1 and 2 refer to two perpendicular directions locally on the sky (i.e. $x$ and $y$ directions on an image of a small patch of sky).  In this formalism, the spin-0 convergence and spin-2 shear are given by \cite{Bacon:2005qr}
\begin{align}\label{eqn:kappa_lensing}
    \kappa &= \frac{1}{2}\partial^*\partial\psi \\
\label{eqn:shear_lensing}
    \gamma &= \gamma_1 + i\gamma_2 = |\gamma|e^{2i\phi} = \frac{1}{2}\partial\partial\psi
\end{align}

\noindent and the spin-1 and spin-3 flexion fields are given by the derivatives of the convergence and shear, respectively:
\begin{align}\label{eqn:F}
    \F &= \F_1 + i\F_2 = |\F|e^{i\phi}  = \frac{1}{2}\partial\partial^*\partial\psi = \partial\kappa, \\
    \label{eqn:G}
    \G &= \G_1 + i\G_2 = |\G|e^{3i\phi} =\frac{1}{2}\partial\partial\partial\psi = \partial\gamma, 
\end{align}

\noindent where $\phi$\footnote{Not to be confused with the coordinate $\phi$ in the Schwarzschild metric} is the polar angle of $\bm{\theta}$. The shear is an anisotropic, elliptical stretching of the source image.  The $\F$-flexion effect is a skewing distortion which manifests as a centroid shift, whereas the $\G$-flexion is a trefoil distortion resulting in a triangularization of the source image.

\subsection{Exotic spacetime metric}

Consider an exotic spacetime metric, as first given by Ref. \cite{Kitamura:2012zy}:
\begin{align}\label{eqn:KNA_metric}
    ds^2 &= -\left(1 - \frac{\varepsilon_1}{r^n}\right)dt^2 + \left(1+\frac{\varepsilon_2}{r^n}\right)dr^2  + r^2d\Omega^2 \nonumber\\
     & \quad + \mathcal{O}(\varepsilon_1^2,\varepsilon_2^2,\varepsilon_1,\varepsilon_2).
\end{align}

\noindent This generalizes the metric such that (i) the spacetime can depend on the inverse distance to the power of $n$ in the weak-field, and (ii) small ``book-keeping" parameters $\varepsilon_1$ and $\varepsilon_2$ are introduced.  Notice that for $n=1$ with negative $\varepsilon_1$ and $\varepsilon_2$, we recover the linearized Schwarzschild metric with a negative Schwarzschild radius and hence negative mass.

With this exotic metric, the deflection angle given in Eq. \eqref{eqn:deflection_sch} becomes \cite{Izumi:2013tya}
\begin{equation}\label{eqn:alpha}
    \hat{\alpha} = \frac{\varepsilon}{b^n}\int_0^{\pi/2}d\phi\cos^n\phi + \mathcal{O}(\varepsilon^2) \equiv \frac{\bar{\varepsilon}}{b^n}
\end{equation}

\noindent where we have defined $\varepsilon \equiv n\varepsilon_1 + \varepsilon_2$, and absorbed the positive-definite integral into $\varepsilon$ such that the sign of $\bar{\varepsilon}$ is the same as $\varepsilon$.  As Ref. \cite{Izumi:2013tya} points out, this deflection angle recovers the Schwarzschild ($n=1$) and Ellis wormhole ($n=2$) cases. We further point out that the SIS ($n=0$) case is also recovered. 

From here, one can obtain an expression for the convergence (in its interpretation as a dimensionless surface mass density).  In terms of the lensing potential, the spin-1 deflection angle is given by $\bm{\alpha} = \partial \psi$.  Therefore, by Eq. \eqref{eqn:kappa_lensing}, $\kappa = \partial^*\bm{\alpha}/2$.  Using this, Eq. \eqref{eqn:alpha}, and the definitions of $\bm{\alpha}$ and $\bm{\theta}$ given in Eq. \eqref{eqn:lens_eqn}, we find that\footnote{Note that this expression includes a coefficient $D_{\rm ds}D_{\rm d}/D_{\rm s}$ not present in Refs. \citep{Kitamura:2012zy, Izumi:2013tya}. This coefficient, with dimensionality $r$ (i.e. distance), is necessary so that convergence is dimensionless: $\bar{\varepsilon}$ has dimensionality $r^n$ [see Eq. \eqref{eqn:KNA_metric}] and impact parameter has dimensionality $r$.}
\begin{equation}\label{eqn:kappa_b}
    \kappa(b) = \frac{D_{\rm ds} D_{\rm d}}{D_{\rm s}}\frac{\bar{\varepsilon}(1-n)}{2}\frac{1}{b^{n+1}}.
\end{equation}

By Eq. \eqref{eqn:alpha}, we see that there exists gravitational attraction on light rays for $\varepsilon > 0$ and repulsion for $\varepsilon < 0$. From Eq. \eqref{eqn:kappa_b}, we note that for $\varepsilon > 0$ and $n>1$ or for $\varepsilon < 0$ and $n <1$, the convergence is negative.  In this context, negative convergence requires that matter (and energy) be exotic. 

For ${\varepsilon} > 0$, there exists a positive root corresponding to the Einstein ring, $\beta = 0$.  For ${\varepsilon} < 0$, on the other hand, there does not exist a positive root corresponding to $\beta = 0$.  This means that there is no Einstein ring for this case, which is to be expected since this case corresponds to light repulsion.  We can still define a typical angular size for this lens, though, as the ``Einstein radius."  In either case, the Einstein radius is defined generally as \cite{Izumi:2013tya}
\begin{equation}\label{eqn:theta_E}
    \theta_E \equiv \left(\frac{|\bar{\varepsilon}| D_{\rm ds}}{D_{\rm s}D_{d}^n} \right)^{1/(n+1)}.
\end{equation}

\section{Flexion in exotic spacetime metrics}{\label{sec:3}}

\subsection{Weak lensing in exotic spacetimes}

Let us restrict ourselves to $\bm{\theta}/\theta_{\rm E} > 0$ (the other image solution occurring for $\bm{\theta}/\theta_{\rm E} < 0$ for $\varepsilon > 0$).  Recognizing that $\bar{\varepsilon} = {\rm sgn}(\varepsilon)|\bar{\varepsilon}|$, where ${\rm sgn}(x)$ is the signum function, and using Eq. \eqref{eqn:theta_E}, we can then write Eq. \eqref{eqn:lens_eqn} as
\begin{equation}
    \bm{\beta} = \bm{\theta} - {\rm sgn}(\varepsilon)\theta_{\rm E}^{n+1} \frac{\bm{\theta}}{\theta^{n+1}}.
\end{equation}

\noindent From here, we can compute the distortion matrix $\mathcal{A}_{ij} \equiv \partial\beta_i/\partial\theta_j$. The elements of this matrix are
\begin{align}\label{eqn:A_ij}
    \mathcal{A}_{11} &= 1 - {\rm sgn}(\varepsilon)\frac{\theta_{\rm E}^{n+1}}{\theta^{n+1}} + {\rm sgn}(\varepsilon)(n+1)\theta_{\rm E}^{n+1}\frac{\theta_1\theta_1}{\theta^{n+3}} \nonumber\\
    \mathcal{A}_{12} &= {\rm sgn}(\varepsilon)(n+1)\theta_{\rm E}^{n+1}\frac{\theta_1\theta_2}{\theta^{n+3}} \nonumber\\
    \mathcal{A}_{21} &= {\rm sgn}(\varepsilon)(n+1)\theta_{\rm E}^{n+1}\frac{\theta_1\theta_2}{\theta^{n+3}} \\
    \mathcal{A}_{22} &= 1 - {\rm sgn}(\varepsilon)\frac{\theta_{\rm E}^{n+1}}{\theta^{n+1}} + {\rm sgn}(\varepsilon)(n+1)\theta_{\rm E}^{n+1}\frac{\theta_2\theta_2}{\theta^{n+3}}. \nonumber
\end{align}

\noindent Reference \cite{Izumi:2013tya} was able to obtain the convergence and shear in the case $\theta_i = (\theta,0)$ and $\beta_i = (\beta,0)$. In this work, we will obtain general expressions for the convergence and both components of the shear.  We do this by working in terms of the second derivatives of the lensing potential. The distortion matrix can be written as
\begin{align}
    \mathcal{A}_{ij} &= \beta_i,_{j} = \delta_{ij} - \psi,_{ij} \nonumber\\
    &=
    \begin{pmatrix}
    1 - \kappa - \gamma_1 & -\gamma_2 \\
    -\gamma_2 & 1-\kappa+\gamma_1
    \end{pmatrix} \\
    &=
    \begin{pmatrix}
    1 - \psi,_{11} & -\psi,_{12} \\
    -\psi,_{12} & 1-\psi,_{22}
    \end{pmatrix}. \nonumber
\end{align}

\noindent Using this with Eq. \eqref{eqn:A_ij}, we see that
\begin{align}\label{eqn:psi_ij}
    \psi,_{11} &= {\rm sgn}(\varepsilon)\frac{\theta_{\rm E}^{n+1}}{\theta^{n+1}} - {\rm sgn}(\varepsilon)(n+1)\theta_{\rm E}^{n+1}\frac{\theta_1^2}{\theta^{n+3}} \nonumber\\
    \psi,_{12} &= -{\rm sgn}(\varepsilon)(n+1)\theta_{\rm E}^{n+1}\frac{\theta_1\theta_2}{\theta^{n+3}} \\
    \psi,_{22} &=  {\rm sgn}(\varepsilon)\frac{\theta_{\rm E}^{n+1}}{\theta^{n+1}} - {\rm sgn}(\varepsilon)(n+1)\theta_{\rm E}^{n+1}\frac{\theta_2^2}{\theta^{n+3}}. \nonumber
\end{align}

\noindent The convergence is therefore given by
\begin{align}\label{eqn:conv_eps_pos}
    \kappa &= \frac{1}{2}(\psi,_{11}+\psi,_{22}) \nonumber\\
     &= {\rm sgn}(\varepsilon)\frac{(1-n)}{2}\frac{\theta_{\rm E}^{n+1}}{\theta^{n+1}}
\end{align}

\noindent and the components of the shear are 
\begin{align}
    \gamma_1 &= \frac{1}{2}(\psi,_{11} - \psi,_{22}) \nonumber\\
     &= -{\rm sgn}(\varepsilon)\frac{(1+n)}{2}\frac{\theta_{\rm E}^{n+1}}{\theta^{n+3}}\left(\theta_1^2 - \theta_2^2\right) \\
    \gamma_2 &= \psi,_{12} \nonumber\\
     &= -{\rm sgn}(\varepsilon)(1+n)\frac{\theta_{\rm E}^{n+1}\theta_1\theta_2}{\theta^{n+3}}.
\end{align}

\noindent The total shear is therefore
\begin{equation}\label{eqn:shear_eps_pos}
    \gamma = -{\rm sgn}(\varepsilon)\frac{(1+n)}{2}\frac{\theta_{\rm E}^{n+1}}{\theta^{n+1}}e^{2i\phi}.
\end{equation}

\noindent Now, the flexion fields can be obtained either by differentiating the convergence and shear directly, or by first obtaining the third derivatives of the lensing potential.  We choose to do the latter for completeness. From Eq. \eqref{eqn:psi_ij}, we obtain
\begin{align}\label{eqn:psi_ijk}
    \psi,_{111} &= -{\rm sgn}(\varepsilon)3(1+n)\theta_{\rm E}^{n+1} \frac{\theta_1}{\theta^{n+3}} \nonumber\\
     &\quad + {\rm sgn}(\varepsilon)(1+n)(3+n)\theta_{\rm E}^{n+1} \frac{\theta_1^3}{\theta^{n+5}} \nonumber\\
    \psi,_{112} &= -{\rm sgn}(\varepsilon)(1+n)\theta_{\rm E}^{n+1} \frac{\theta_2}{\theta^{n+3}} \nonumber\\
     &\quad + {\rm sgn}(\varepsilon)(1+n)(3+n)\theta_{\rm E}^{n+1} \frac{\theta_1^2\theta_2}{\theta^{n+5}} \nonumber\\
    \psi,_{122} &= -{\rm sgn}(\varepsilon)(1+n)\theta_{\rm E}^{n+1} \frac{\theta_1}{\theta^{n+3}} \nonumber\\
    &\quad + {\rm sgn}(\varepsilon)(1+n)(3+n)\theta_{\rm E}^{n+1} \frac{\theta_1\theta_2^2}{\theta^{n+5}} \nonumber\\
    \psi,_{222} &= -{\rm sgn}(\varepsilon)3(1+n)\theta_{\rm E}^{n+1} \frac{\theta_2}{\theta^{n+3}} \nonumber\\
    &\quad +{\rm sgn}(\varepsilon)(1+n)(3+n)\theta_{\rm E}^{n+1} \frac{\theta_2^3}{\theta^{n+5}}.
\end{align}

\noindent From this, the components of the $\F$-flexion are given by
\begin{align}
    \F_1 &= \frac{1}{2}(\psi,_{111} + \psi,_{122}) \nonumber\\
     &= -{\rm sgn}(\varepsilon)\frac{(1-n^2)}{2}\theta_{\rm E}^{n+1}\frac{\theta_1}{\theta^{n+3}} \\
    \F_2 &= \frac{1}{2}(\psi,_{112} + \psi,_{222}) \nonumber\\
     &= -{\rm sgn}(\varepsilon)\frac{(1-n^2)}{2}\theta_{\rm E}^{n+1}\frac{\theta_2}{\theta^{n+3}}    
\end{align}

\noindent such that the total $\F$-flexion is
\begin{equation}\label{eqn:F_eps_pos}
    \F = -{\rm sgn}(\varepsilon)\frac{(1-n^2)}{2}\frac{\theta_{\rm E}^{n+1}}{\theta^{n+2}}e^{i\phi},
\end{equation}

\noindent and the components of the $\G$-flexion are given by
\begin{align}
    \G_1 &= \frac{1}{2}(\psi,_{111} - 3\psi,_{122}) \nonumber\\
     &= {\rm sgn}(\varepsilon)\frac{(1+n)(3+n)}{2}\theta_{\rm E}^{n+1}\frac{\theta_1^3-3\theta_1\theta_2^2}{\theta^{n+5}} \\
    \G_2 &= \frac{1}{2}(3\psi,_{112} - \psi,_{222}) \nonumber\\
     &= {\rm sgn}(\varepsilon)\frac{(1+n)(3+n)}{2}\theta_{\rm E}^{n+1}\frac{3\theta_1^2\theta_2-\theta_2^3}{\theta^{n+5}}    
\end{align}

\noindent such that the total $\G$-flexion is
\begin{equation}\label{eqn:G_eps_pos}
    \G = {\rm sgn}(\varepsilon)\frac{(1+n)(3+n)}{2}\frac{\theta_{\rm E}^{n+1}}{\theta^{n+2}}e^{3i\phi}
.\end{equation}

As pointed out by Ref. \cite{Izumi:2013tya}, the expression for convergence given by Eq. \eqref{eqn:conv_eps_pos} is consistent with the Schwarzschild lens for $\varepsilon > 0$ and $n=1$.  We point out that this is also true of the shear, given by Eq. \eqref{eqn:shear_eps_pos}, and the flexions, given by Eqs. \eqref{eqn:F_eps_pos} and \eqref{eqn:G_eps_pos}.  Additionally, we note that all four lensing fields recover the SIS lens for $\varepsilon > 0$ and $n=0$. One will also immediately notice that the lensing fields in the $\varepsilon < 0$ case are the negatives of the $\varepsilon > 0$ case. 

\subsection{Discussion of shear and flexion behavior}

\subsubsection{Lensing field signatures from ordinary matter} 

Let us first discuss the behavior of convergence, shear, and flexion in nonexotic, typical weak lensing situations.  As an illustrative example, consider the SIS lens ($\varepsilon > 0$ and $n=0$). This type of lens corresponds to an ordinary, positive mass, for which $\kappa>0$.  We can study the behavior of the shear and flexion lensing fields by considering the simple example of a source (background) galaxy located at polar angle $\phi = 0$ around the lens.  Then, the direction of the field is encoded in the sign of the lensing field amplitude.  Around such a lens, there exists ``tangential" alignment of galaxy ellipticities, such that the shear given by Eq. \eqref{eqn:shear_eps_pos} $\gamma < 0$ (again, for $\phi = 0$). $\F$-flexion, which has the spin properties of a vector, points radially toward the lens, such that $\F < 0$ in Eq. \eqref{eqn:F_eps_pos}. $\G$-flexion oscillates around the lens as a spin-3 quantity; however, its behavior could also be described as a type of radial alignment, but where $\G > 0$ in Eq. \eqref{eqn:G_eps_pos}. 

This is the behavior of lensing fields in the presence of some ordinary positive mass, and therefore it also describes the picture of galaxy-galaxy lensing.  Furthermore, this lensing signature is also found in lensing by cosmological large-scale structure.  In cosmology, there exists matter density perturbations relative to some mean density in the Universe.  There are regions of mass overdensity (a mass peak) and regions of mass underdesnity (a mass trough). The mass overdensity can be modeled as, for example, an SIS lens.  In the presence of a mass overdensity, there is tangential shear alignment and radial flexion alignment of background galaxies. Lensing fields display the opposite behavior when a mass underdensity is a lens: there is tangential anti-alignment\footnote{Tangential anti-alignment is referred to as ``cross"-alignment in the literature} of shear: $\gamma > 0$ (again, for $\phi = 0$) and anti-radial alignment for flexion: $\F > 0$ and $\G < 0$.

This is all to say that, when we simultaneously observe tangential alignment of shear and radial alignment of flexion, we expect there to be a mass peak (e.g. a positive mass lens or some local mass overdensity) and when anti-tangential alignment of shear and anti-radial alignment of flexion is observed, we expect the lens to be some local mass underdensity. 

\subsubsection{Lensing field signatures from exotic matter} 

\begin{figure}[htb!]
  \centerline{\includegraphics[width=\linewidth]{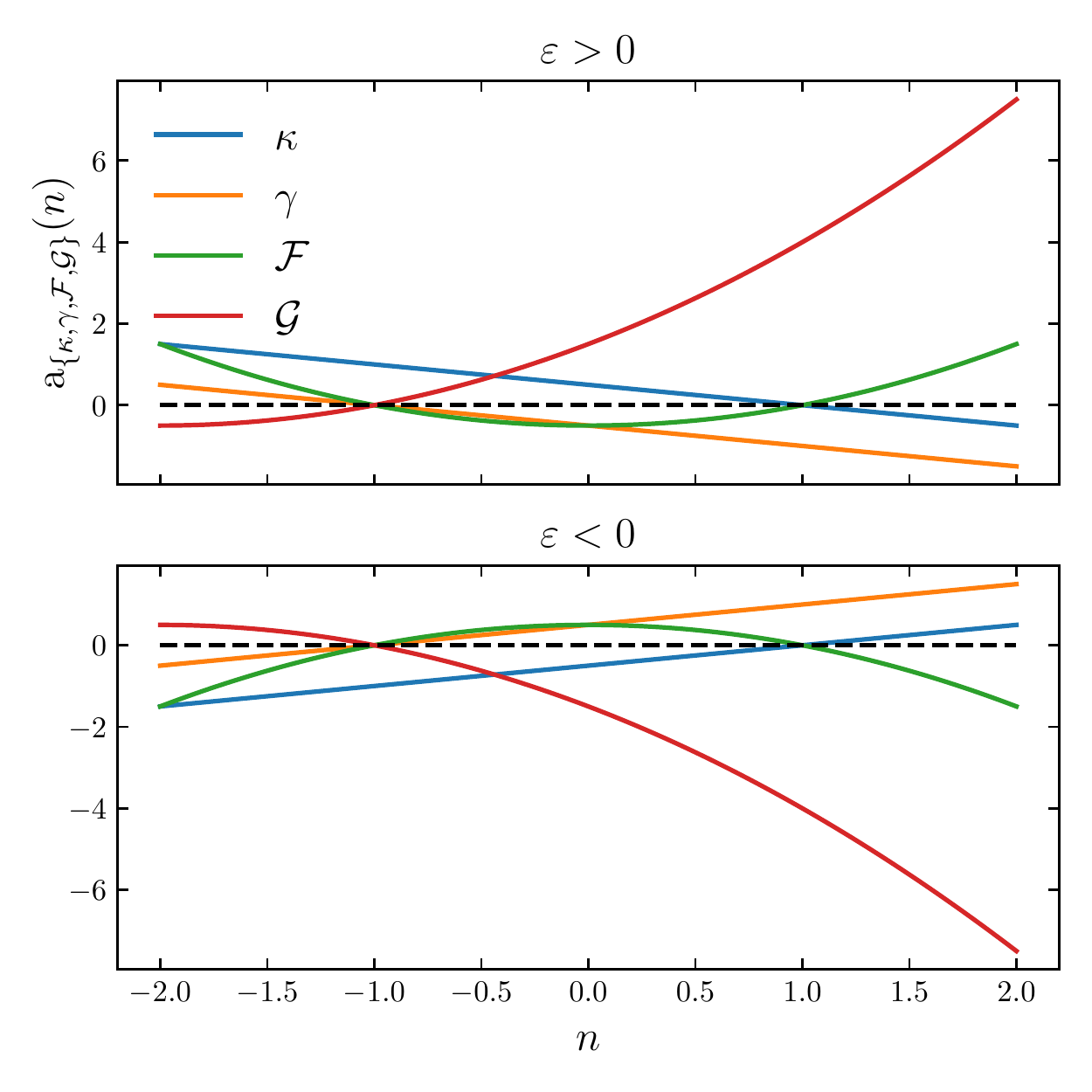}}
  \caption{Here we plot the amplitudes as a function of $n$ for convergence, shear, $\F$-flexion, and $\G$-flexion for the cases $\varepsilon > 0$ (top) and $\varepsilon < 0$ (bottom). The signs of each amplitude for shear and flexion indicate alignment around a lens (for $\phi = 0)$, whereas the sign on the amplitude for convergence indicates positive or negative convergence. Here we see how the $\F$-flexion is coupled to the convergence and the $\G$-flexion to the shear for $n > 1$. Specifically, we note that the $\F$-flexion behaves opposite to that of the shear in the presence of an $n > 1$ lens.}
\label{fg:amp}
\end{figure}

Let us now turn our attention to weak lensing in the exotic spacetime metric. Figure \ref{fg:amp} shows the magnitude and algebraic sign of the amplitudes of the various lensing fields for the $\varepsilon > 0$ and $\varepsilon < 0$ cases, respectively.  

First, we consider the $\varepsilon > 0$ case, where we know from the deflection angle in Eq. \eqref{eqn:alpha} that there exists a gravitational attraction on light rays from the lens. If and only if $n > -1$, tangential alignment of shear exists.  Interestingly, for $\varepsilon > 0$ and $n>1$, the convergence is negative, despite there being a tangential shear.  As discussed earlier, this negative convergence corresponds to exotic matter (and energy).  This was pointed out in Ref. \cite{Izumi:2013tya}; however, in this work, we have additional information from the flexion fields.  Here, radial alignment of $\F$-flexion only exists for $\varepsilon > 0$ and $-1 < n < 1$. For $n>1$, there is an anti-radial $\F$-flexion alignment despite the fact that there is a tangential alignment of shear.  $\G$-flexion, on the other hand, is radially aligned for $n > -1$.  This is all to say that there is a stark difference in the behavior of the lensing fields for $n>1$.  Whereas shear and $\G$-flexion would indicate the presence of a mass-overdense lens that pulls on light rays, the convergence and $\F$-flexion behave as if there is a mass-underdense lens.  This is a consequence of the fact that $\F$-flexion is the derivative of convergence and $\G$-flexion is the derivative of shear. Therefore, $\F$-flexion responds locally to convergence, and $\G$-flexion responds nonlocally to the shear. 

Next, we consider the $\varepsilon < 0$ case, in which there exists a gravitational repulsion on light rays from the lens. Here, we observe precisely the opposite behavior as in the $\varepsilon > 0$ case. For $n>1$, the convergence and $\F$-flexion behave as if there is a mass-overdense lens, whereas the shear and $\G$-flexion act as if there is a mass-underdense lens.  

\begin{figure}
  \centerline{\includegraphics[width=\linewidth]{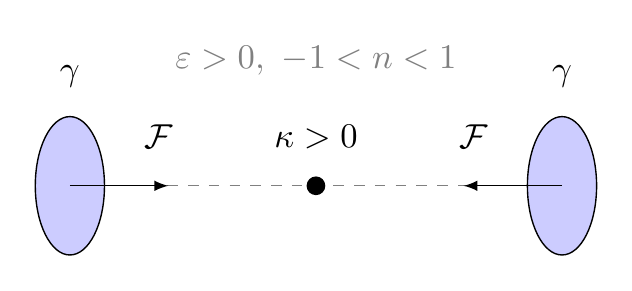}}
  \vspace{-2em}
\centerline{\includegraphics[width=\linewidth]{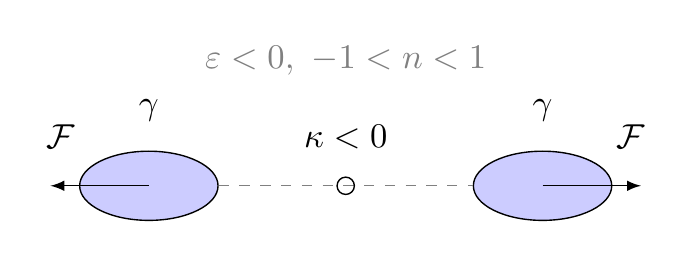}}
  \vspace{-2em}
  \centerline{\includegraphics[width=\linewidth]{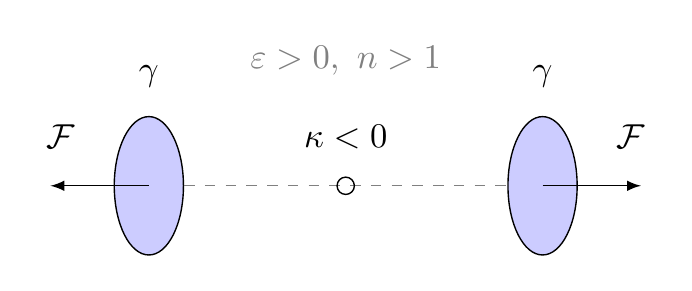}}
  \caption{The behavior of shear and $\F$-flexion for  two source (background) galaxies on opposite sides ($\phi = 0$ and $\phi = \pi$ radians) of three different lenses. \textit{Top}: a typical scenario of lensing by a nonexotic mass defined by $\varepsilon > 0$ and $-1<n<1$ (we exclude the $n=1$ Schwarzschild case where convergence and $\F$-flexion vanish).  The shear has a tangential alignment around the lens and the $\F$-flexion points radially toward the lens (radial alignment).  \textit{Middle}: lensing by some exotic object ($\kappa < 0)$ defined by $\varepsilon < 0$ and $-1<n<1$.  Here, shear has tangential anti-alignment and $\F$-flexion points radially outward (radial anti-alignment).  \textit{Bottom}: lensing by some exotic object ($\kappa <0$) defined by $\varepsilon > 0$ and $n>1$.  Here, shear has tangential alignment and $\F$-flexion points radially outward. }
\label{fg:cartoon}
\end{figure}

In Fig. \ref{fg:cartoon}, we show a cartoon of the behavior of shear and $\F$-flexion for three different cases.  In each case, there are two source galaxies on opposite sides of a lens (one at $\phi = 0$ and the other at $\phi = \pi$ radians). The top panel recovers the typical case of lensing by a nonexotic object such as, for example, the SIS lens.  In the middle case, we see the shear and $\F$-flexion responding to a negative convergence with $\varepsilon < 0$ and  $-1<n<1$.  In this formalism, this is considered an exotic lens.  In the context of cosmology; however, this could be interpreted as the lensing fields responding to a local mass underdensity, such as a cosmic void \cite{Izumi:2013tya}. Finally, the case on the bottom recovers that of an Ellis-wormhole type metric. Here, there is a negative convergence for the exotic object.  As discussed earlier, while the shear is tangentially aligned, behaving as if there is an overdense lens, the flexion behaves as if there is an underdense lens, responding to the convergence.

We can also remark on the behavior of the lensing fields in the presence of a negative-mass compact object. A compact object can be described by the Schwarzschild metric ($n=1$). For the case of a positive mass ($\varepsilon > 0$), shear is tangentially aligned, $\G$-flexion is radially aligned, but the convergence and $\F$-flexion, interestingly, both vanish.  Flexion can add additional information when looking for negative-mass compact objects ($\varepsilon < 0)$: one would expect the following lensing signature: tangential anti-alignment of shear, radial anti-alignment of $\G$-flexion, and $\F=0$.     

\subsubsection{Flexion to the rescue}

In the work presented in Ref. \cite{Izumi:2013tya}, it is difficult to distinguish between various exotic lenses.  Since convergence is not a directly observable quantity, one is relying entirely on shear. For example, two vastly different lenses -- a lens with a positive convergence versus one with a negative convergence -- are both capable of exhibiting identical directional behavior for shear.  This can only be disambiguated by examining the relative strengths of the shear signals; however, this method could be difficult for, e.g. an $n \rightarrow 1^{-}$ negative convergence versus an $n \rightarrow 1^{+}$ positive convergence.  

There is also another difficulty with relying only on shear, which comes from the fact that, unlike in e.g. galaxy-galaxy lensing, exotic lenses such as the Ellis wormhole may be completely invisible.  This means that the lens position is unknown.  Shear responds rather weakly to substructure, so using tangential alignment in order to locate the position of a lens, while possible, may not be ideal. 

Both of these problems can be ameliorated with the addition of flexion.  $\F$-flexion in particular is locally connected to the convergence, such that its directional information can distinguish between negative and positive convergences that cause identical shear directional patterns.  Additionally, $\F$-flexion responds strongly to substructure, and so it could be more easily used to identify an unknown lens position.  $\G$-flexion could be useful as a systematics check between shear and $\F$-flexion.  While it is complementary to $\F$-flexion in terms of strength, it should follow the directional behavior of the shear. 

In Ref. \cite{Izumi:2013tya}, it can also prove difficult to distinguish between ordinary and exotic lenses. The Ellis wormhole ($\varepsilon > 0$ and $n = 2$) is an illustrative example of this. If relying on directional information alone, the Ellis wormhole creates a tangentially aligned shear pattern that resembles that of a typical nonexotic positive-mass lens.  When we add flexion, however, the Ellis wormhole produces a unique lensing signature: tangential alignment of shear, anti-radial alignment of $\F$-flexion, and radial alignment of $\G$-flexion. To the best of our knowledge, only an exotic lens is capable of producing this type of lensing signature.  Therefore, when combined with shear, flexion can be used to uniquely associate particular lensing signatures to exotic objects. 

\section{Cosmic flexion in modified gravity}{\label{sec:4}}

In the standard $\Lambda$CDM model of cosmology, the field equations of GR describe the relationship between spacetime geometry and the matter-energy content of the Universe governed by gravity. The Friedmann–Lema{\^i}tre–Robertson–Walker (FLRW) metric describes a homogeneous and isotropic Universe.  To quantify gravitational lensing in a cosmological context, however, it is necessary to consider scalar perturbations in the FLRW metric.  In the conformal Newtonian gauge, the line element of such a metric is given by \cite{1995ApJ...455....7M}
\begin{equation}
    ds^2 = a^2(\tau)\left[\left(1+\frac{2\Psi}{c^2}\right)c^2d\tau^2  - \left(1-\frac{2\Phi}{c^2}\right)dl^2 \right]
\end{equation}

\noindent where $\tau$ is the conformal time, $a$ is the scale factor, and $dl^2 = d\chi^2 + f_K^2(\chi)d\Omega^2$, where $f_K(\chi)$ is the comoving angular distance, which is simply equal to $\chi$ for a flat Universe, in which case $d l^2 = \delta_{ij} dx^i dx^j$. The two Bardeen potentials, $\Psi(\bm{x},\tau)$ and $\Phi(\bm{x},\tau)$ are considered to describe weak fields, $\Psi,\Phi \ll c^2$. In GR, the two Bardeen potentials are equal to each other:
\begin{equation}
    \Phi_{\rm N} = \Phi = \Psi,
\end{equation}

\noindent where $\Phi_{\rm N}$ is the Newtonian gravitational potential defined via the Poisson equation.  In modified gravity, these potentials need not be equivalent.  

The local deflection of light rays -- propagating along null geodesics -- relative to unperturbed ones, depends on the light travel time obtained from the metric:
\begin{equation}
    \frac{d\tau}{dl} \approx \frac{1}{c}\left[1-\frac{1}{c^2}(\Phi+\Psi)\right].
\end{equation}

\noindent Using this to obtain the deflection $d\alpha$, integrating over comoving distance, and using the lens equation, one obtains the cosmological effective convergence (obtained for GR in Ref. \cite{Bartelmann:1999yn}), 
\begin{align}
    \kappa_{\rm eff}(\bm{\theta}, \chi) &= \frac{1}{2c^2}\int_0^{\chi}d\chi'\frac{f_K(\chi-\chi')f_K(\chi')}{f_K(\chi)}\nonumber\\
    & \quad\quad\quad\quad\times \nabla^2[\Psi+\Phi](f_K(\chi')\bm{\theta},\chi')
\end{align}

\noindent where $\bm{\theta}$ is the angular position on the sky, $\chi$ is the comoving distance (along the line of sight), and the Laplacian is given by $\nabla^2 =\partial^2/\partial x_i\partial x_i + \partial^2/\partial \chi^2$, where summation over $i$ is implied, and $x$ are physical distances perpendicular to the line of sight. 

In GR, the linearized Einstein equations relate the metric perturbations (the Bardeen potentials) to the perturbations of the cosmological fluid.  These include the matter density contrast $\delta \equiv \delta\rho/\overline{\rho}$, the pressure perturbation $\delta p$, the divergence of the fluid velocity $\theta$, and the stress or anisotropic pressure $\sigma$.  Computing the Einstein equations is most easily done in Fourier space, where we exchange spatial derivatives with powers of $i\bm{k}$, where $k$ is the comoving wavenumber.  A combination of the $0-0$ and $0-i$ equations yields the generalized Poisson equation \cite{1995ApJ...455....7M, 2011A&A...530A..68T}:
\begin{equation}\label{eqn:Poisson}
    -k^2\tilde{\Phi} = 4\pi G a^2 \overline{\rho} \tilde{\Delta} = \frac{3}{2}\Omega_{m,0}H_0^2a^{-1}\tilde{\Delta}
\end{equation}

\noindent where $H_0$ is the Hubble constant, $\Omega_{m,0}$ is the matter density parameter at present epoch, 
\begin{equation}
    \tilde{\Delta}(\bm{k},a) = \tilde{\delta}(\bm{k},a) + \frac{H(a)(1+\theta)}{k^2}
\end{equation}

\noindent is the comoving density perturbation, $H(a)$ is the Hubble parameter, $w=\overline{p}/\overline{\rho}$ is the equation of state parameter, and the second equality in Eq. \eqref{eqn:Poisson} is written for matter only. We can parametrize deviations from GR through use of the mass-screening phenomenological post-GR function $\tilde{Q}(\bm{k},a)$, replacing Newton's gravitational constant by an effective function
\begin{equation}
    G_{\rm eff} = G\tilde{Q} \implies -k^2\tilde{\Phi} = 4\pi G \tilde{Q} a^2 \overline{\rho} \tilde{\Delta}.
\end{equation}

\noindent From the $i-j$ Einstein equation, one obtains
\begin{equation}
    k^2(\tilde{\Phi}-\tilde{\Psi}) = 12\pi G a^2\overline{\rho}(1+w)\sigma.
\end{equation}

\noindent Here, we can pursue a further deviation from GR which quantifies the difference in the Bardeen potentials, $\Psi - \Phi$, through use of the gravitational slip phenomenological post-GR function, $\tilde{\eta}(\bm{k},a)$:
\begin{equation}
    \tilde{\Psi} = (1+\tilde{\eta})\tilde{\Phi} \implies k^2(\tilde{\Phi}-\tilde{\Psi}) = 4\pi G\tilde{Q}\tilde{\eta} a^2 \overline{\rho}\tilde{\Delta}.
\end{equation}

\noindent  From this, we obtain
\begin{align}\label{eqn:Poisson_MG}
    -k^2(\tilde{\Phi}+\tilde{\Psi}) &= 8\pi G\tilde{Q}\left(1+\frac{\tilde{\eta}}{2}\right) a^2 \overline{\rho} \tilde{\Delta} = 8\pi G\tilde{\Sigma} a^2 \overline{\rho} \tilde{\Delta} \\
    -k^2\tilde{\Psi} &= 4\pi G Q(1+\tilde{\eta})a^2\overline{\rho}\tilde{\Delta} = 4\pi G \tilde{\Gamma}a^2\overline{\rho}\tilde{\Delta} 
\end{align}

\noindent where we have defined $\tilde{\Sigma}(\bm{k},a) = \tilde{Q}(\bm{k},a)(1+\tilde{\eta}(\bm{k},a)/2)$ and $\tilde{\Gamma}(\bm{k},a) = \tilde{Q}(\bm{k},a)(1+\tilde{\eta}(\bm{k},a))$.  These two derived post-GR functions are commonly used in cosmic shear studies.  In general, these functions can depend on both cosmic time and scale \cite{2011A&A...530A..68T}.  

\subsection{Case 1: Scale-independent post-GR functions}

The two derived post-GR functions have been taken to be scale-independent for various cosmic shear studies (e.g. see Ref. \cite{2011A&A...530A..68T}).  In this case, we have the following simplification for Eq. \eqref{eqn:Poisson_MG}:
\begin{equation}
    -k^2\left[\tilde{\Phi}+\tilde{\Psi}\right](\bm{k},a) = 3\Omega_{m,0}H_0^2a^{-1} \Sigma(a) \tilde{\Delta}(\bm{k},a).
\end{equation}

\noindent Taking the Fourier transform of this, the cosmological effective convergence can be written \cite{2011A&A...530A..68T} 
\begin{align}
    \kappa_{\rm eff}(\bm{\theta}, \chi) &= \frac{3}{2}\Omega_{m,0}\left(\frac{H_0}{c}\right)^2\int_0^{\chi}d\chi'\frac{f_K(\chi-\chi')f_K(\chi')}{f_K(\chi)a(\chi')}\nonumber\\
    & \quad\quad\quad\quad\times \Sigma(a(\chi'))\Delta(f_K(\chi')\bm{\theta},\chi').
\end{align}

\noindent This gives the effective convergence for a fixed source redshift corresponding to a comoving distance $\chi$.  When the sources are distributed in comoving distance, the cosmological effective convergence needs to be averaged over the (normalized) source distribution $n(\chi)$.  This is to say that $\kappa_{\rm eff}(\theta) = \int_0^{\chi_{H}} d\chi n(\chi)\kappa_{\rm eff}(\bm{\theta}, \chi)$, where $\chi_{H}$ is the horizon distance obtained for infinite redshift.  By introducing the lensing efficiency function, 
\begin{equation}\label{eqn:q}
q(\chi) = \frac{3}{2}\Omega_{m,0} \left(\frac{H_0}{c}\right)^2\frac{f_K(\chi)}{a(\chi)} \int_{\chi}^{\chi_H}d\chi' n(\chi') \frac{f_K(\chi' - \chi)}{f_K(\chi')},
\end{equation}

\noindent and rearranging integration limits, we obtain
\begin{equation}
    \kappa_{\rm eff}(\bm{\theta}) = \int_0^{\chi_{H}} d\chi q(\chi)\Sigma(a(\chi))\Delta(f_K(\chi)\bm{\theta},\chi).
\end{equation}

\noindent Limber's equation/approximation states that for two quantities, $g_a$ and $g_b$ of the form $g_a = \int_0^{\chi_H}d\chi h_a(\chi)X(f_K(\chi)\bm{\theta},\chi)$ where $X$ is some field, e.g. the density contrast, the cross-power spectrum of $g_a$ and $g_b$ is \citep{Bartelmann:1999yn, 1953ApJ...117..134L, LoVerde:2008re}
\begin{equation}
    \Pow_{ab}(\ell) = \int_0^{\chi_H} d\chi \frac{h_a(\chi)h_b(\chi)}{f_K^2(\chi)}\Pow_X\left(k=\frac{\ell+1/2}{f_K(\chi)},\chi\right).
\end{equation}

\noindent where $\ell$ is the angular wavenumber. If one sets $h_a = h_b = q(\chi)\Sigma(a(\chi))$, we obtain the convergence power spectrum
\begin{equation}
    \Pow_{\kappa}(\ell) = \int_0^{\chi_H} d\chi \frac{q^2(\chi)\Sigma^2(a(\chi))}{f_K^2(\chi)}\Pow_\Delta^{\rm MG}\left(k=\frac{\ell+1/2}{f_K(\chi)},\chi\right)
\end{equation}

\noindent where $\Pow_\Delta^{\rm MG}(k,z) \neq \Pow_\Delta^{\rm GR}(k,z)$ is the (nonlinear) matter power spectrum in modified gravity.  It is clear that deviations to GR modify the cosmic shear power spectrum amplitude via $\Sigma(a)$.  They further impact the power spectrum, though, via a modification of the matter power spectrum.  This is because the evolution of the density contrast is modified via Eq. \eqref{eqn:Poisson}, leading to a different cosmic evolution than in GR \cite{2009PhRvD..80b3532D}. 

Next, we wish to obtain the cosmic flexion power spectrum \cite{Bacon:2005qr, AGB}.  Making use of the definition of flexion given by Eq \eqref{eqn:F}, i.e.
\begin{equation}
    \F_i = \partial_i\kappa = \frac{\partial}{\partial\theta_i}\kappa = f_K(\chi)\frac{\partial}{\partial x_i}\kappa,
\end{equation}

\noindent the cosmological effective flexion can be written as
\begin{align}
    \F_{\rm eff}(\bm{\theta}, \chi) &= \frac{3}{2}\Omega_{m,0}\left(\frac{H_0}{c}\right)^2\int_0^{\chi}d\chi'\frac{f_K(\chi-\chi')f_K(\chi')}{f_K(\chi)a(\chi')}\nonumber\\
    & \quad\quad\quad\quad\times \Sigma(a(\chi'))f_K(\chi')\Delta'(f_K(\chi')\bm{\theta},\chi')
\end{align}

\noindent where $\Delta'$ is the transverse gradient of the density contrast.  We therefore obtain $\F_{\rm eff}(\bm{\theta}) = \int_0^{\chi_{H}} d\chi q(\chi)\Delta'(f_K(\chi)\bm{\theta},\chi)$.  This time, we set $h_a = h_b = f_K(\chi)q(\chi)\Sigma(a(\chi))$ and obtain, via Limber's equation, 
\begin{align}
        \Pow_{\F}(\ell) &= \int_0^{\chi_H} d\chi q^2(\chi)\Sigma^2(a(\chi))\Pow_{\Delta'}^{\rm MG}\left(k=\frac{\ell+1/2}{f_K(\chi)},\chi\right) \nonumber\\
         &= \ell^2\Pow_{\kappa}(\ell)
\end{align}

\noindent where we have noted that $|X'|^2 = |X|^2k_ik^i$ and hence (taking $k=(\ell+1/2)/f_K(\chi) \approx \ell/f_K(\chi)$) \cite{Bacon:2005qr}
\begin{equation}
    \Pow_{X'}\left(\frac{\ell}{f_K(\chi),\chi}\right) = \Pow_{X}\left(\frac{\ell}{f_K(\chi),\chi}\right)\frac{\ell^2}{f_K^2(\chi)}.
\end{equation}

In addition to the cosmic flexion power spectrum, we can also obtain the convergence-flexion cross-spectrum \citep{Bacon:2005qr, AGB}.  We again use Limber's equation, but this time we work in terms of $\Pow_\Delta^{\rm MG}$ rather than $\Pow_{\Delta'}^{\rm MG}$.  We set $h_{\kappa} = q(\chi)\Sigma(a(\chi))$ and $h_{\F} = q(\chi)\Sigma(a(\chi))\ell$ to obtain
\begin{align}
    \Pow_{\kappa\F}(\ell) &= \ell\Pow_{\kappa}(\ell).
\end{align}

\noindent We note that, owing to the fact that shear and convergence statistics are the same, i.e. $\Pow_{\gamma}(\ell) = \Pow_{\kappa}(\ell)$ \cite{Kilbinger:2014cea}, so too [because of the relations in Eq. \eqref{eqn:G}] are the $\F$- and $\G$-flexion power spectra,
\begin{equation}
    \Pow_{\G}(\ell) = \Pow_{\F}(\ell),
\end{equation}

\noindent and similarly, 
\begin{equation}
        \Pow_{\kappa\G}(\ell) = \Pow_{\kappa\F}(\ell).
\end{equation}

\subsection{Case 2: Scale-dependent post-GR functions}

In general the two derived post-GR functions are functions of scale.  In Fourier space, 
\begin{equation}
    -k^2\left[\tilde{\Phi}+\tilde{\Psi}\right](\bm{k},a) = 3\Omega_{m,0}H_0^2a^{-1} \tilde{\Sigma}(\bm{k},a) \tilde{\Delta}(\bm{k},a).
\end{equation}

\noindent We define the following quantity:
\begin{equation}\label{eqn:Delta_Sigma}
    \tilde{\Delta}_{\Sigma}(\bm{k},a) \equiv \tilde{\Sigma}(\bm{k},a) \tilde{\Delta}(\bm{k},a),
\end{equation}

\noindent and obtain the cosmological effective convergence 
\begin{align}
    \kappa_{\rm eff}(\bm{\theta}, \chi) &= \frac{3}{2}\Omega_{m,0}\left(\frac{H_0}{c}\right)^2\int_0^{\chi}d\chi'\frac{f_K(\chi-\chi')f_K(\chi')}{f_K(\chi)a(\chi')}\nonumber\\
    & \quad\quad\quad\quad\times \Delta_{\Sigma}(f_K(\chi')\bm{\theta},\chi').
\end{align}

\noindent Again using Limber's equation, we obtain the convergence power spectrum
\begin{equation}
        \Pow_{\kappa}(\ell) = \int_0^{\chi_H} d\chi \frac{q^2(\chi)}{f_K^2(\chi)}\Pow_{\Delta_\Sigma}^{\rm MG}\left(k=\frac{\ell+1/2}{f_K(\chi)},\chi\right),
\end{equation}

\noindent where $\Pow_{\Delta_\Sigma}^{\rm MG}(k,z)$ is defined in Fourier space via Eq. \eqref{eqn:Delta_Sigma}. Following the same steps as before, the cosmological effective convergence is given by
\begin{align}
    \F_{\rm eff}(\bm{\theta}, \chi) &= \frac{3}{2}\Omega_{m,0}\left(\frac{H_0}{c}\right)^2\int_0^{\chi}d\chi'\frac{f_K(\chi-\chi')f_K(\chi')}{f_K(\chi)a(\chi')}\nonumber\\
    & \quad\quad\quad\quad\times f_K(\chi')\Delta_\Sigma'(f_K(\chi')\bm{\theta},\chi').
\end{align}

\noindent We immediately note that the cosmological effective convergence depends on the transverse gradient of $\Delta_\Sigma$, and therefore it is a probe of the derivative of the derived post-GR function $\Sigma$.  This positions flexion as a unique probe of modified gravity, allowing the measurement of $\Sigma'$ alongside $\Sigma$. 

Finally, we find that $\Pow_{\F}(\ell) = \Pow_{\G}(\ell) = \ell^2\Pow_{\kappa}(\ell)$ and $\Pow_{\kappa\F}(\ell) = \Pow_{\kappa\G}(\ell) =  \ell\Pow_{\kappa}(\ell)$, as before. Cosmic shear-shear, flexion-flexion, and shear-flexion correlations probe different scales (see Ref. \cite{AGB} for a detailed discussion).  While cosmic shear has a broad window function for power at the scale of arcminutes, cosmic flexion peaks at the arcsecond scale, with shear-flexion peaking intermediate to these two signals.
With the use of cosmic flexion in addition to cosmic shear, there exists the opportunity to probe the behavior of modified gravity across a wide range of cosmic scales and times.      

\section{Conclusions}{\label{sec:con}}

In this work we have considered the weak gravitational flexion that is induced by exotic lenses, such as the Ellis wormhole, through use of an exotic spacetime metric.  We have also reported a more generalized expression for the weak gravitational shear.  We show that the analytic equations for convergence, shear, and the flexions in this exotic spacetime recover familiar nonexotic lenses such as the Schwarzschild lens and the SIS lens.  We find that flexion can provide valuable information about exotic lenses when used in addition to shear.  In particular, the directional information from $\F$-flexion can be used to distinguish between positive and negative convergences, and can provide unique lensing signatures for objects such as the Ellis wormhole, whereas the directional information from shear alone cannot.  

We also consider cosmic flexion in the context of modified gravity.  We find that the cosmological effective flexion depends on the transverse spatial derivative of the derived phenomenological post-GR function $\Sigma$, positioning flexion as a unique probe of parametric modified gravity.  Additionally, we are able to construct cosmic flexion-flexion and shear-flexion power spectra, which probe different scales than cosmic shear, allowing for further exploration of deviations from GR, particularly in the case of scale-dependent post-GR functions. 


\begin{acknowledgments}
E. J. Arena would like to thank David M. Goldberg, David J. Bacon, and Jacob Shpiece for useful conversations about this work.  E. J. Arena also thanks the anonymous referee for very helpful suggestions regarding the uniqueness of the different exotic lensing signatures.
\end{acknowledgments}

\bibliography{bibliography}

\end{document}